\documentclass[twocolumn,showpacs,preprintnumbers,amsmath,amssymb,prl]{revtex4}

\usepackage{graphicx}
\usepackage{dcolumn}
\usepackage{bm}

\begin{document}


\title{Jamming proteins with slipknots and their free energy landscape}                 

\author{Joanna I. Su{\l}kowska$^{1}$, Piotr Su{\l}kowski$^{\,2,3,4}$ and Jos{\'e} N. Onuchic$^1$}

\affiliation{
$^1$ Center for Theoretical Biological Physics, University of California San Diego, Gilman Drive 9500, La Jolla 92037, \\
$^2$ Physikalisches Institute and Bethe Center for Theoretical Physics,
Universit{\"a}t Bonn, Nussallee 12, 53115 Bonn, Germany \\
$^3$ California Institute of Technology, Pasadena, CA 92215,\\
$^4$ Institute for Nuclear Studies, Ho{\.z}a 69, 00-681 Warsaw, Poland \\}



\begin{abstract}

Theoretical studies of stretching proteins with slipknots reveal a surprising growth
of their unfolding times when the stretching force crosses an intermediate threshold.
This behavior arises as a consequence of the existence of alternative  unfolding routes 
that are dominant at different force ranges.
Responsible for longer unfolding times at higher forces is the existence 
of an intermediate, metastable configuration where the slipknot is jammed.
Simulations are performed with a coarsed grained model with further quantification  using a refined description of the geometry of the slipknots.
The simulation data is used  to determine the free energy landscape (FEL) of the
protein, which supports recent analytical predictions.

\end{abstract}

\pacs{87.15.ap, 87.14.E-, 87.15.La, 82.37.Gk, 87.10.+e}

\maketitle


The large increase in determining new protein structures has led to the
discovery of several proteins with complicated topology. This new fact has arised the question
if their energy landscape and the folding mechanism is similar to typical proteins.
One class of such proteins includes knotted proteins which comprise around 1\% of all
structures deposited in the PDB database \cite{Mansfield,Virnau}. 
A related class of proteins contains more subtle geometric configurations called slipknots
 \cite{Yeates_2007b,Taylor_2007}. 
Recent theoretical studies using structure-based models (where native contacts are dominant) suggest
that slipknot-like conformations act like intermediates during the folding of knotted proteins \cite{Sulkowska_2008f}. 
This entire new mechanism is consistent with energy landscape theory (FEL) and the funnel concept \cite{Onuchic_1992,Onuchic2004}.
It was shown that the slipknot formation reduces the  topological barrier. 
Complementing regular folding studies, additional information about the landscape was obtained by 
mechanical manipulation of the knotted protein with atomic force microscopy \cite{R4} both
 experimentally in \cite{Alam_2002,Rief_2008} 
and theoretically in \cite{Sulkowska_2008,Sulkowska_2008e,Dziubiela_2009}.
For example, \cite{Sulkowska_2008} it has been showen that unfolding proceeds via a series of jumps
between various metastable conformations, a mechanism opposite to the smooth unfolding in knotted homopolymers.

Motivated by these early results, we now propose a unified picture for the mechanical unfolding 
of proteins with slipknots. In this Letter   this question is addressed by explaining the role of topological barriers 
along their mechanical unfolding pathways. 
Supported by our previous results that knotted proteins can still have a minimally frustrated funnel-like energy landscape,
 structure-based theoretical coarse-grained models are used
\cite{Sulkowska_2007b} to analyze the behavior of a slipknot protein under stretching. 
Studies are performed for the $\alpha/\beta$ 
class protein thymidine kinase (PDB code: 1e2i \cite{1e2i}).

\begin{figure}[htb]
\begin{center}
\includegraphics[width=0.36\textwidth]{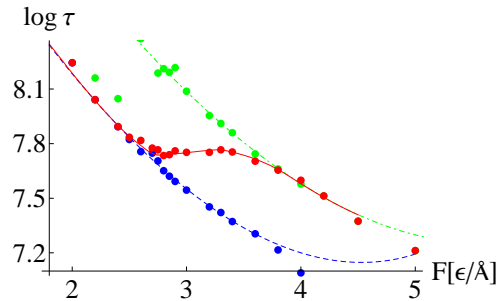}
\caption{Dependence of the unfolding times $\tau$ on the stretching force $F$ for 1e2i (solid line, in red).
In this Letter we describe this mechanism as  a superposition of two unfolding pathways: 
I for small forces (dashed (lower) line, in blue), and II for intermidiate and 
large forces (dashed-dotted (upper) line, green).} \label{razem}
\end{center}
\end{figure}

Most of our analysis is based on stretching
simulations under constant force \cite{Onuchic1999}. The crucial signature for this process
is the overall unfolding time from the beginning of the stretching
until the protein fully unfolds. 
Normally one expects that the transition between the native and the unfolded
basins to be limited by overcoming the free energy barrier, which gets effectively reduced
upon an application of a stretching force. The rate by which this barrier is reduced depends 
on the distance between the unfolded basin and the top of the barrier measured along the stretching coordinate $x$.
This idea was first developed in the phenomenological model 
of Bell \cite{Bell78}, which states that the unfolding time $\tau$ 
decreases exponentially with applied stretching force $F$ as $\tau(F) = \tau_0 e^{-\frac{F x}{k_B T}} $.
A refined analysis performed in ref.~\cite{Dudko-PRL} revealed
that this dependence is more complicated but still monotonically decreasing.  

The unfolding times for 1e2i measured in our simulations are shown as the red curve in Fig. \ref{razem}.
 In contrast to the above expectations, 
increasing the force in the range $3$-$3.5 \epsilon/$\AA$\ $ surprisingly results in a larger stability of the protein. 
$\epsilon$ is the typical effective energy of tertiary native contacts that is consistent with 
the value $\epsilon/$\AA $\simeq$71 pN derived in \cite{Sulkowska_2007b}.
 A solution for this paradox is accomplished by realizing
that unfolding is dominated by two  distinct, alternative routes that are dominant at different force regimes.
A routing switch occurs when  threshold is crossed between weak and intermediate forces.
At higher forces, mechanical unfolding is dominated by a route that involves a 
jammed slipknot. This jamming gives rise to the unexpected dependence of unfolding time on applied force. 
Characterizing this mechanism is the central goal of this Letter.
 
\begin{figure}[htb]
\begin{center}
\includegraphics[width=0.20\textwidth]{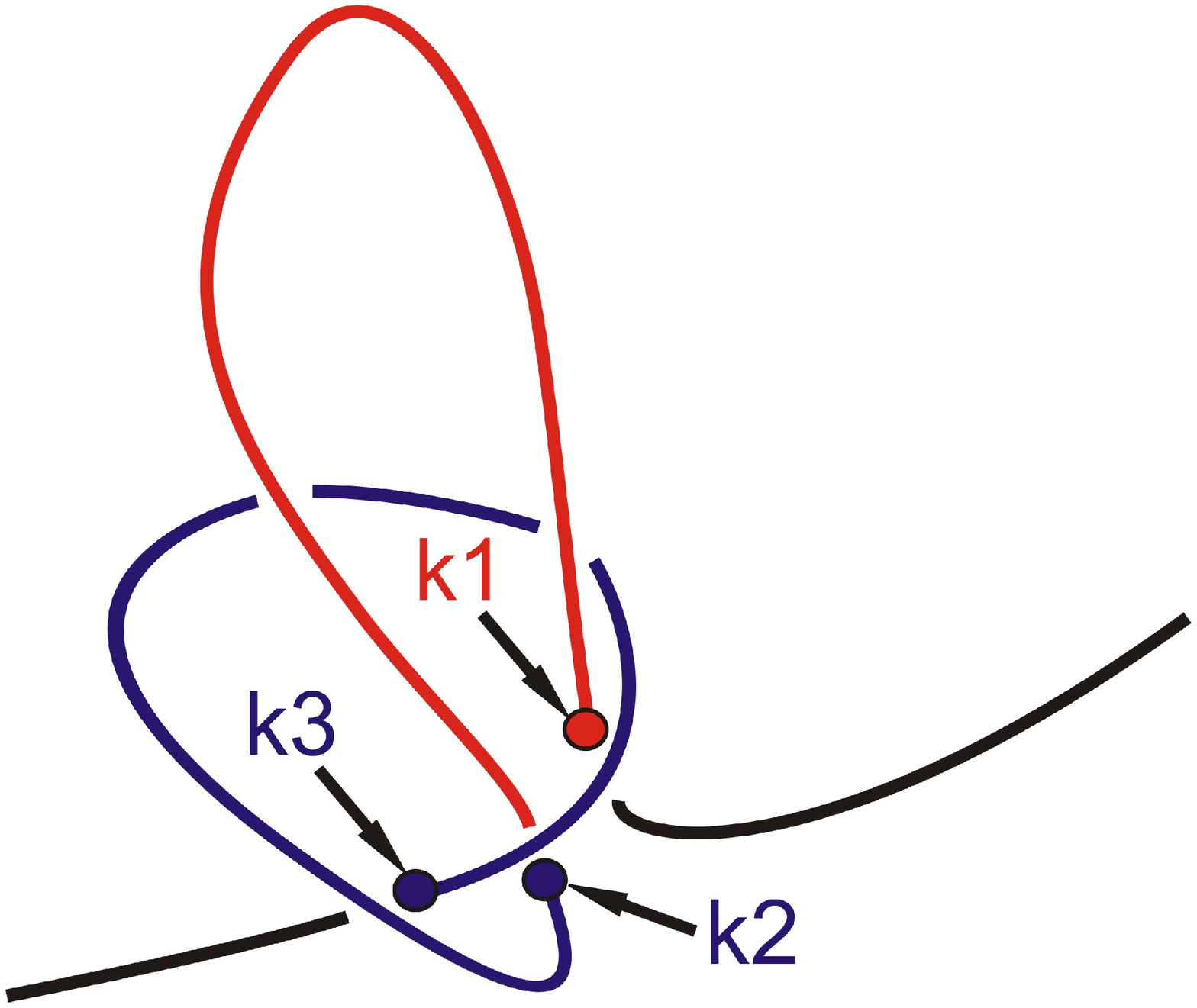}
\includegraphics[width=0.22\textwidth]{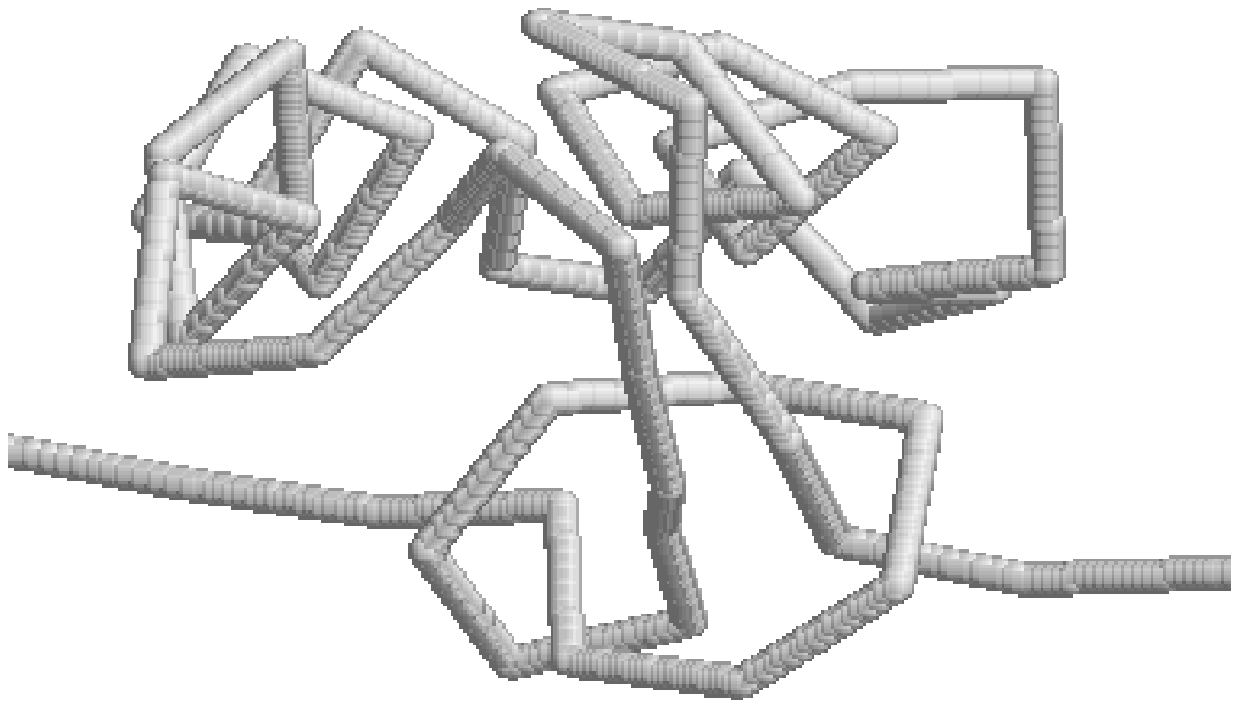}
\caption{A slipknot (left) consists of a \emph{threaded loop} ($k_1-k_2$, in red) which is partialy threaded
through a \emph{knotting loop} ($k_2-k_3$, in blue). 
An example of a protein configuration with
a tightened slipknot is shown in the right panel.} \label{slipknot}
\end{center}
\end{figure}

\begin{figure}[htb]
\begin{center}
\includegraphics[width=0.38\textwidth]{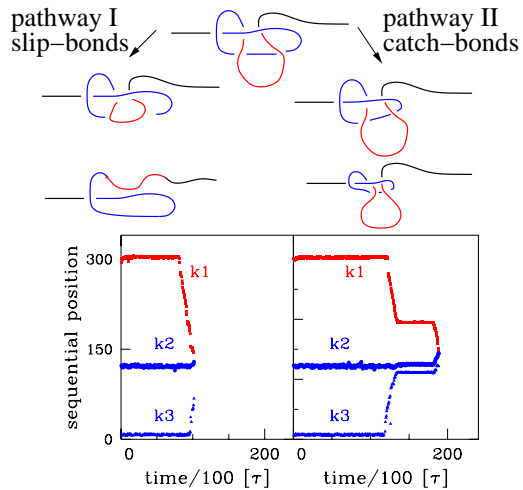}
\caption{The behavior of the slipknot during stretching (top) is determined by the relative
behavior of its two loops, encoded in the time dependence of
$k_1,k_2$ and $k_3$ (bottom). If the threaded loop shrinks faster than 
the knotting loop, $k_1$ merges with $k_2$ (bottom left) and the slipknot untightens 
(pathway I, top left). If the knotting loop shrinks faster, $k_2$ approaches $k_3$ (bottom right, $\simeq 14000\tau$) and
the slipknot gets temporarily tightened (pathway II, top right).
This is a metastable state which can eventually untie further stretching , with $k_1$ finally merging 
with $k_2$ (bottom right, $\simeq 19000\tau$). Kinetic studies were performed slightly above folding temperature 
using overdamped Langevin dynamics with typical folding times of $10000\tau$.      
} \label{endpoints}
\end{center}
\end{figure}

To describe the evolution of a slipknot quantitatively requires a refined description. 
A slipknot is characterized by the three points shown in Fig. \ref{slipknot}. 
The first point $k_1$ is determined by eliminating 
amino acids consecutively from one terminus until the knot configuration is reached (which can be detected e.g. 
by applying the KMT algorithm \cite{km1}). The two additional
points, $k_2$ and $k_3$, correspond to the ends of this knot. 
In the native state the protein 1e2i contains a slipknot with $k_1=10, k_2=128, k_3=298$.
These three points divide the slipknot
into two loops, which are called the \emph{knotting loop} and the \emph{threaded
 loop}. The former one is the loop of the trefoil knot and the latter
one is threaded through the knotting loop. Unfolding of the slipknot upon stretching
depends on the relative shrinking velocity of these two loops (see Fig. \ref{endpoints}).
When the threaded loop shrinks faster than the knotting loop, the slipknot unties.
In the opposite case the slipknot gets (temporarily) tightened or jammed, resulting in a metastable
state associated to a local minimum in the protein's FEL. 
Upon further stretching, this configuration eventually also unties. The evolution
of both loops of the slipknot is encoded in the time dependence of the points
$k_1, k_2, k_3$, see Fig. \ref{endpoints}.

Before discussing the stretching of 1e2i, we explain why a
slipknot formed by a uniformly elastic polymer should smoothly unfold under stretching. 
To simplify the discussion we approximate the threaded and knotting loops  by circles of size $R_t$ and $R_k$.
These two loops shrink during stretching and, when the threaded one eventually 
vanishes, the slipknot gets untied.
If both loops have similar sizes, the slipknot is very unstable and unties immediately.
When the threaded loop is much larger than the knotting one, $R_t> > R_k$, 
 untightening can be explained as follows. 
The elastic energy associated to local bending is proportional to
the square of the curvature. If the loop is approximated  by a circle of radius $R$,
then its local curvature is constant and equals $R^{-1}$. The total elastic energy is
$\oint ds R^{-2} \sim R^{-1}$ \cite{Landau}. From the assumption $R_t> > R_k$
we conclude that upon stretching it is energetically favorable to decrease $R_t$ rather than $R_k$.
This happens until both radii become equal and then, just as above, the slipknot gets very unstable and untightens.
In this discussion we have not yet taken into account that when a slipknot is stretched some parts of 
a chain slide along each other.
This effect could be incorporated by including  the friction generated by the sliding
\cite{elasticknots}.  But in the slipknot the sliding
region associated with the knotting loop is much longer than the region associated
to the threaded loop. Thus this effect results in a faster tightening of the threaded
rather than the knotting loop, facilitating even more the untightening of the slipknot.

The above argument should apply to slipknots in biomolecules because they are characterized by a persistence length
that in principle is simply related to their elasticity \cite{cos1}.
For DNA this effect is described by worm-like-chain models (WLC) \cite{cos2}
and it has been confirmed experimentally.
Although WLC models are too simple to describe the protein general behavior, they are useful in
some limited applications. Thus at first sight one might expect that slipknots in proteins 
should smoothly untie upon stretching.
Proteins, however, are much more complicated than DNA or uniformly elastic polymers. 
The presence of stabilizing native tertiary contacts leads to a jumping character during stretching 
\cite{Sulkowska_2008}. In addition their bending energy is not uniform along the chain due to the heterogeneity
of the amino-acid sequence. As a consequence it turns out that the intuition obtained through the above analysis of polymers
or WLC models is misleading. 

\begin{figure}[htb]
\begin{center}
\includegraphics[width=0.38\textwidth]{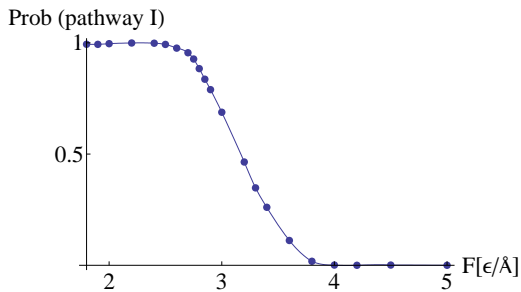}
\caption{Dependence on the applied stretching force of the probability 
of choosing pathway I rather than II (see Fig. \ref{endpoints}). 
This varying probability leads to the complicated dependence of the total unfolding 
time on the stretching force observed in Fig. \ref{razem}.
} \label{probI}
\end{center}
\end{figure}

Our analysis of the evolution of the endpoints $k_1,k_2,k_3$ (Fig. \ref{endpoints}, bottom) reveals that
for various stretching forces unfolding proceeds along two distinct pathways (Fig. \ref{endpoints}, top). 
In pathway I the slipknot smoothly unties, which is observed 
for relatively weak forces. At intermediate forces pathway II starts to dominate and
the knotting loop can shrink tightly before the threaded one vanishes.
In this regime the protein gets temporarily jammed (Fig. \ref{endpoints}, right), leading to much longer unfolding  times (catch pathway). 
The probability of choosing pathway I at different forces is shown in Fig. \ref{probI}. 
This pathway competition explains the nontrivial total unfolding time dependence observed in Fig. \ref{razem}.

The two different pathways I and II  arise from
completely different unfolding mechanisms.
Pathway I starts and continues mostly from the C-terminal side, along 
16$\alpha$, 15$\beta$, 14$\alpha$, 13$\beta$, 12(helices bundle), 11$\alpha$
(here the number denotes a consecutive secondary structure as counted from N-terminal, 
and $\alpha$ or $\beta$ specifies whether this is a helix or a $\beta$-sheet; for more details about the structure of 1e2i see the PDB). 
This is followed by unfolding of helices 11$\alpha$, 10$\alpha$ that allows breaking
of the contacts inside the $\beta$-sheet created by the N-terminal, with unfolding proceeding also from the N-terminal. 
Pathway II also starts from the C-terminal but rapidly 
(as soon as helix 15 is unfolded) switches to the N-terminal. In this case, differently from pathway I,  
the $\beta$-sheet from the N-terminal unfolds even before 13$\beta$. 
These scenarios indicate that the pathway I should be dominant
at weak forces since they are not sufficient to break
the $\beta$-sheet during first steps of unfolding.
The jammed pathway is typical only if stretching forces are sufficiently strong for
unfolding to proceed from the two terminals of the protein. 

A similar phenomenon was firstly proposed in ref.~\cite{Dembo_1988} and referred to as catch-bonds.
Experimental evidence suggesting this mechanism was first observed for adhesion complexes \cite{Zhu_2003,Vogel}.
Using AFM, at large forces the ligand-receptor pair becomes entangled and therefore expands the unfolding time.
A theoretical description of this mechanism was given in ref.~\cite{Ajdari,Zhu_2004,Thirumalai_2005}.



The kinetic data can also  be used to determine the associated  free
energy landscape (FEL) \cite{Onuchic_1992}. 
In an initial simplification we associate the barriers along the stretching coordinate as the the kinetic bottlenecks during the mechanical unfolding event.
Generalizing Bell's model,  a recent description of two-state mechanical unfolding in the presence of a single transition barrier
has been developed in \cite{Dudko-PRL}, with the rate equation
\begin{equation}
\tau(F) = \tau_0 \Big(1-\frac{\nu F x^{\dagger}}{\Delta G} \Big)^{1-1/\nu}  e^{-\frac{\Delta G}{k_B T} \big(1-(1-\nu F x^{\dagger} / \Delta G)^{1/\nu}   \big)},
\label{DHSz}
\end{equation}
where $\nu$ encodes the shape of the barrier. 
Here $x^{\dagger}$ denotes the distance between the barrier 
and the unfolded basin (in a first approximation it can be regarded
as $F$ independent) and lies on the reaction coordinate along the AFM pulling direction.
It can be experimentally determined by measuring how
the stretching force modulates the unfolding times $\tau$. The height of the barrier is denoted by $\Delta G$.
Fig. \ref{razem} (unfolding times are given by solid red line) 
shows that this single barrier theory is not sufficient for the full range of forces. 
As described before, in the higher force regime,
additional basins have to be included in the energy landscape. 
Models with several metastable basins have been called multi-state FEL models \cite{Rief-EMBO2005}. 
Evidence supporting the need of multi-states FEL was confirmed by AFM experiments in different systems
\cite{Clarke-Nature2003,Bustamante-Science2005}. 
 
\begin{figure}[htb]
\begin{center}
\includegraphics[width=0.26\textwidth]{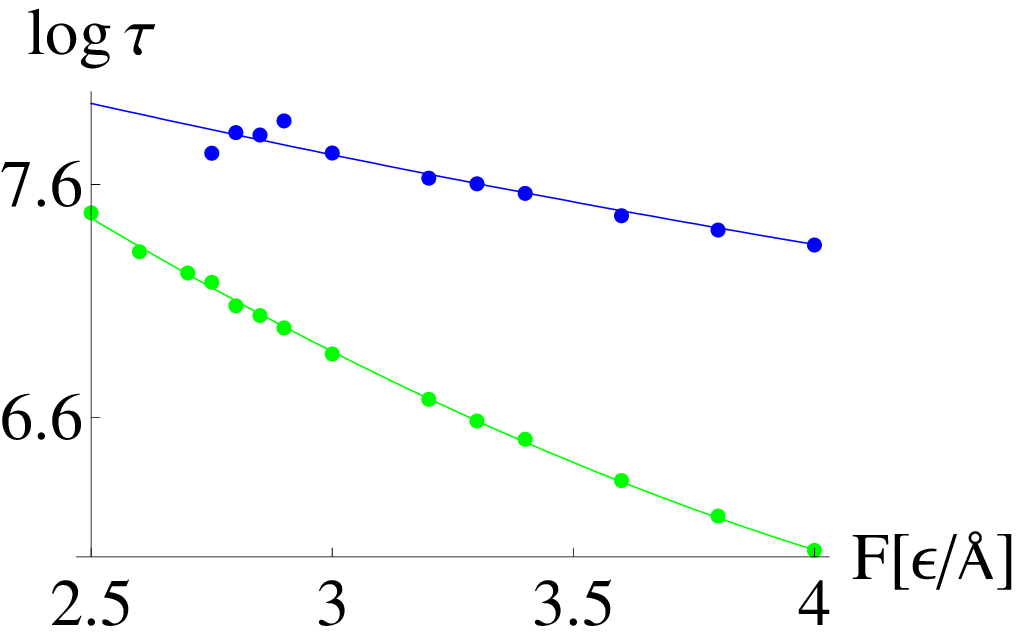} 
\includegraphics[width=0.18\textwidth]{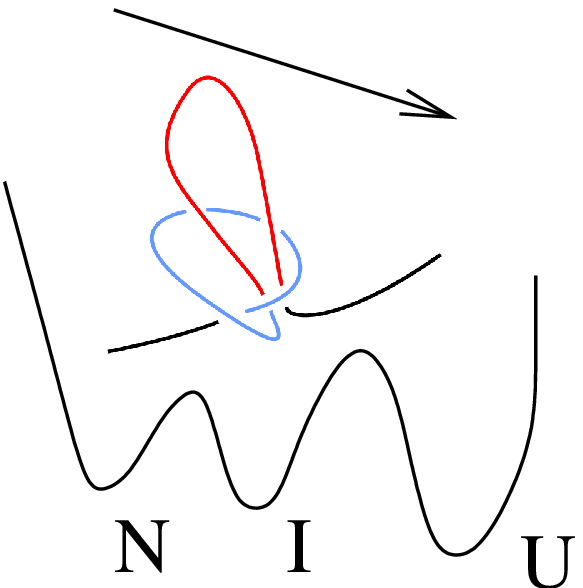}
\caption{Pathway II with two barriers. Left:
dependence of the unfolding time on the applied force 
with the data and the fit to the formula (\ref{DHSz}) for the first maximum (lower, in green) 
and for the second maximum (upper, in blue). 
Right: schematic free energy landscape for this pathway, with jammed slipknot 
in a minimum between two barriers.} \label{barriers-II}
\end{center}
\end{figure}

To construct a multi-state FEL that incorporates two unfolding pathways I and II 
we use  a linear combination of 
eq.~(\ref{DHSz})-like  expressions with different shapes and barrier heights.
Each one of them essentially accounts for the distinct barrier along a relevant unfolding route.
Fitting the stretching data  to eq.~(\ref{DHSz}) with a cusp-like  $\nu=1/2$ approximation 
(another possibility $\nu=2/3$ for the cubic potential in general leads to similar results \cite{Dudko-PRL})
determines accurately the location and the height of the potential barriers.
Pathway II involves two barriers: first until the moment of creation of the 
intermediate which is followed  the untieing event.
They are characterized by $(x_1,\Delta G_{1})$ and $(x_2,\Delta G_{2})$ 
arising respectively from the lower and upper fits in Fig. \ref{barriers-II} (left). 
The superposition of these two fits gives the overall mean unfolding 
time for pathway II (dotted-dashed curve in green in Fig. \ref{razem}). 
For the ordinary slipknot unfolding (pathway I),
the results $x_{I}$ and $G_{I}$ arise from the dashed blue curve in Fig. \ref{razem}. This analysis leads to the results
$$
x_1 = 2.3 \frac{k_B T \textrm{\AA}}{\epsilon}, 
\quad x_2 = 0.7 \frac{k_B T \textrm{\AA}}{\epsilon}, 
\quad x_{I} = 1.4 \frac{k_B T \textrm{\AA}}{\epsilon}, 
$$
$$
\Delta G_1 = 8.0 k_B T, 
\quad \Delta G_2 = 4.2 k_B T, 
\quad \Delta G_{I} = 4.7 k_B T.
$$
We conclude that the free energy landscape consists of two ``valleys". The force-dependent probability of choosing
one of the valleys during stretching depends on the details of the protein structure. It is determined from our
simulations as shown in Fig. \ref{probI}. Using these probability values and the parameters above for
 $x$ and $\Delta G$, we can accurately represent the simulation data using a linear combination of 
 equations of the form (\ref{DHSz}). 
This agreement supports our analytical analysis and generalizes eq. (\ref{DHSz})
for the full of range forces. In addition it demonstrates that structure-based models sufficiently capture the major geometrical 
properties of a slipknotted protein. 
A schematic representation of the free energy landscape for pathway II is shown in Fig. \ref{barriers-II} (right).

Summarizing, we have analyzed the process of tightening of the slipknot in protein
1e2i and determined the corresponding free energy landscape. 
Its main feature is the presence of a metastable configuration with a tightened slipknot,
which is observed for sufficiently large pulling forces. 
This phenomenon does not exist for uniformly elastic polymers.
In this Letter we concentrated on protein 1e2i but similar behavior has also been observed  
for other proteins with slipknots, e.g. 1p6x.
Our results provide testable predictions that can now be verified by
AFM stretching experiments.

\medskip 

We appreciate useful comments of O. Dudko.
The work of J.S. was supported by the Center for Theoretical Biological Physics 
sponsored by the NSF (Grant PHY-0822283) with additional support from NSF-MCB-0543906.
P.S. acknowledges the support of Humboldt Fellowship, DOE grant DE-FG03-92ER40701FG-02, 
Marie-Curie IOF Fellowship, and Foundation for Polish Science.



\end{document}